\documentclass[5p,twocolumn,times,number]{elsarticle}
 
\usepackage{hyperref}
\usepackage{lineno}
\modulolinenumbers[5] 
\hypersetup{
    pdfauthor=author
}
\usepackage{amsmath}
\usepackage{amssymb}
\usepackage[english]{babel}
\usepackage[utf8]{inputenc}
\usepackage{csquotes}
\usepackage{color}
\usepackage{textcomp}
\usepackage{xspace}
\usepackage[font=small, format=hang, labelfont={bf}]{caption}
\allowdisplaybreaks
\usepackage{listings}
\usepackage{siunitx}
\DeclareSIUnit \parsec {pc}
\usepackage{sidecap}
\usepackage{wrapfig}
\usepackage{float}
\usepackage{subfig}
\usepackage{psfrag}
\usepackage{tikz}
\usetikzlibrary{quotes,angles}
\usepackage{shellesc}
\usepackage{booktabs}
\usepackage{longtable}
\usepackage{multirow}
 
\journal{Nuclear Instruments and Methods in Physics Research A}

\begin{document}

\begin{frontmatter}

    \title{Development of a SiPM Pixel Prototype for the Large-Sized Telescope of the Cherenkov Telescope Array}

    \author[unito,infnto,mpik]{D. Depaoli\corref{mycorrespondingauthor}}
    \ead{davide.depaoli@mpi-hd.mpg.de}
    \author[unito,infnto]{A. Chiavassa}
    \author[infnpd]{D. Corti}
    \author[infnto]{F. Di Pierro}
    \author[infnpd,unipd]{M. Mariotti}
    \author[infnpd,unipd]{R. Rando}
    \cortext[mycorrespondingauthor]{Corresponding author}
    %
    \address[unito]{Università degli Studi di Torino, Dipartimento di Fisica, Via Pietro Giuria 1, 10125 Torino, Italy}
    \address[infnto]{Istituto Nazionale di Fisica Nucleare, Sezione di Torino, Via Pietro Giuria 1, 10125 Torino, Italy}
    \address[mpik]{Max Planck Institute for Nuclear Physics, Saupfercheckweg 1, 69117 Heidelberg, Germany}
    \address[infnpd]{Istituto Nazionale di Fisica Nucleare, Sezione di Padova, Via Francesco Marzolo 8, 35131 Padova, Italy}
    \address[unipd]{Università degli Studi di Padova, Dipartimento di Fisica e Astronomia ``Galileo Galilei'', Via Francesco Marzolo 8, 35131 Padova, Italy}
    %
    \date{}
    %
    \begin{abstract}
        The Cherenkov Telescope Array (CTA) will be the next generation ground-based gamma-ray observatory. CTA consists of different telescope types of which the largest ones (Large-Sized Telescopes, LSTs) cover the lower energy range, between \SI{20}{\giga\electronvolt} and \SI{200}{\giga\electronvolt}.
        The first LST is currently being commissioned at the Roque de los Muchachos Observatory, La Palma, Canary Islands. Its camera has \num{1855}  photomultipliers (PMTs) with \num{1.5} inch cathodes.
        Silicon Photomultipliers (SiPMs) are increasingly becoming valid alternatives to PMTs also in gamma-ray astronomy.
        In the context of the LST project, there is an effort to study a novel Advanced Camera, equipped with SiPMs and a completely redesigned electronics based on a fully digital approach.
        To study and develop solutions on the sensors of these camera, we
        built a prototype camera module with a fully re-designed pre-amplifying stage and sensor bias control while re-using the digitizing and triggering stages of the existing LST camera module. 
        We report on the design choices made to achieve the highest performance in terms of timing and charge resolution and the laboratory measurements validating those choices.

    \end{abstract}

    \begin{keyword}
        Silicon Photomultipliers \sep 
        Gamma-Ray Astronomy \sep
        Imaging Atmospheric Cherenkov Telescope \sep
        Front-end electronics \sep
        Electronics \sep
        Low light detectors
    \end{keyword}

\end{frontmatter}


\section{Introduction}
Very-high-energy (VHE) gamma rays are excellent probes to study and
understand the origin of cosmic rays and the mechanisms of particle acceleration in extreme environments.
Imaging Atmospheric Cherenkov Telescopes (IACTs) have evolved into the most sensitive instruments in VHE gamma-ray astrophysics. Multiple telescopes acquire stereoscopic images of extensive air showers (EAS) induced by gamma rays in the atmosphere by capturing Cherenkov light. \par
The Cherenkov Telescope Array (CTA) will be the most sensitive ground-based gamma-rays observatory in the energy range between \SI{20}{\giga\electronvolt} and \SI{300}{\tera\electronvolt}. It will be built at two sites, one in each hemisphere, to achieve full-sky coverage.
At each site, IACTs of different sizes ensure coverage of the entire energy range. The largest IACTs (Large-Sized Telescopes, LSTs) will focus on the lower energy range
(\SI{20}{\giga\electronvolt} - \SI{200}{\giga\electronvolt}).
Currently, the first LST is being commissioned at the northern site, the Roque de Los Muchachos Observatory, La Palma, Canary Islands (Spain) \cite{LST_ICRC_2021}. \par
The LST's \qty{23}{m} diameter segmented light-collection surface focuses the Cherenkov light on a
camera with \num{1855} photomultipliers (PMTs) with \qty{1.5}{inch} diameter photocathodes.
Due to their high photon detection efficiency and tolerance to high illumination levels, Silicon Photomultipliers (SiPMs) are viable alternatives to PMTs in IACTs. In CTA, both small-sized  and dual-mirror medium-sized telescopes will use SiPMs \cite{SST_ICRC_2021, SCT_ICRC_2021}.
\par
We are working towards replacing PMTs with SiPMs also in LSTs.  The idea is not only to replace the PMTs with SiPMs, but to completely redesign the electronics with a fully digital approach \cite{Heller_ICRC_2021}.
\par
As a proof-of-concept, we built a SiPM-based module prototype, starting with the current camera configuration and electronics.
The high night sky background (NSB) rate (expected to be around \qty[]{1.5}{\giga\hertz} per pixel in dark condition for our prototype) puts constraints on how long the electrical signals must be. Indeed, the longer the signals,
the higher the trigger threshold must be in order to not be dominated by accidental triggers due to the pile up of overlapping NSB signals.
To reach thresholds that are competitive with PMTs, it was our goal to design the front-end electronics such that the processed SiPM signals are as short as the PMT signals in the current camera.
\par
A similar effort is ongoing in the Major Atmospheric Gamma Imaging Cherenkov (MAGIC) Collaboration, where SiPM modules are installed  for direct comparison with PMTs in one of the two telescopes \cite{MAGIC_SiPM, MAGIC_SiPM_2022}.
%
\section{SiPM-based prototype for the LST camera}
\begin{figure}[!tp]
    \centering
    \includegraphics[width=\columnwidth]{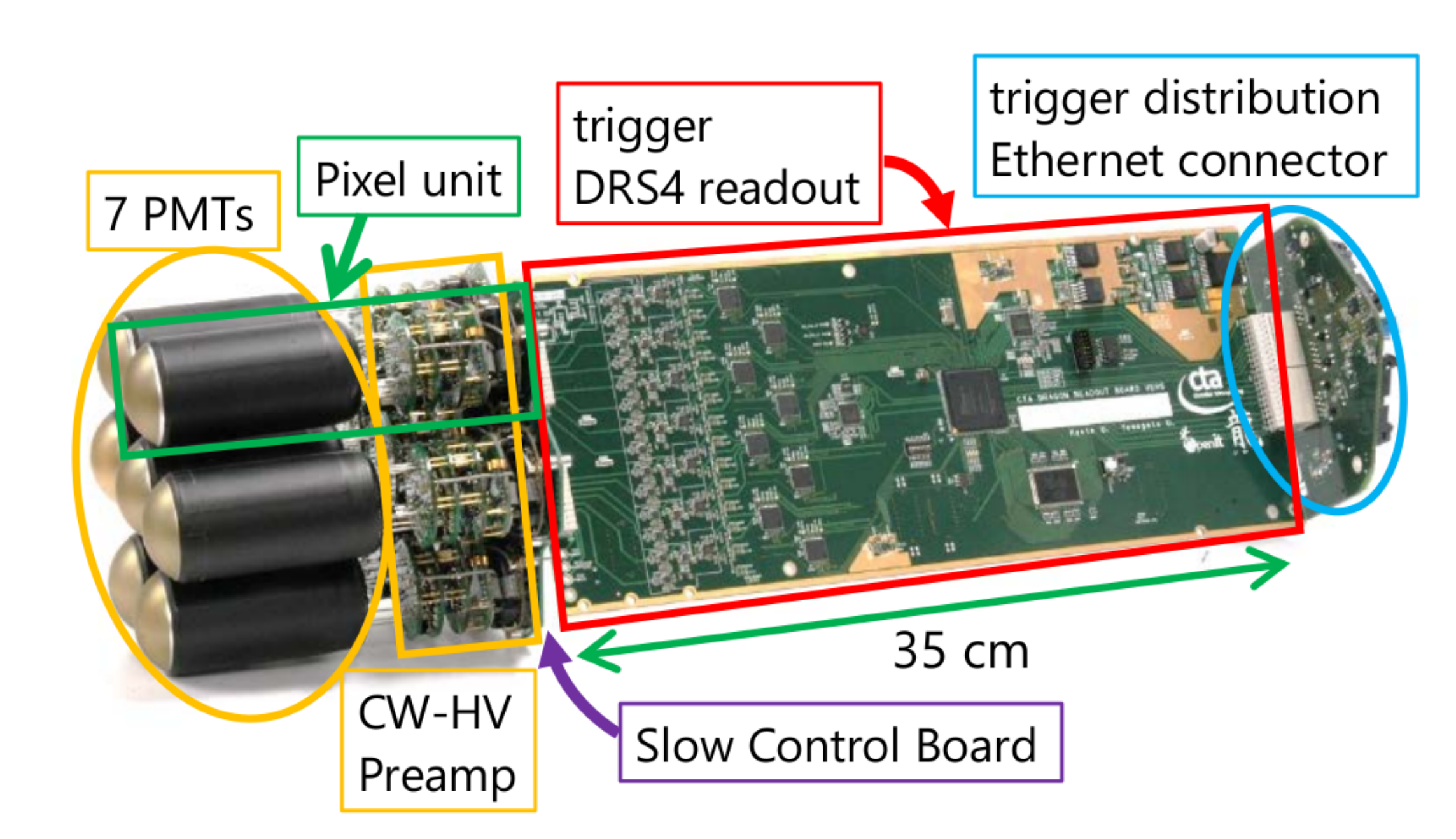}
    \caption{LST camera module \cite{Masu_DevPMT}. Only the Pixel Unit (consisting of the Cockcroft-Walton generator, the Preamplifier and the PMT) is replaced with a SiPM pixel.}
    \label{fig:LST_PMT_01}
\end{figure}

The aim of this work is to build a pixel prototype for a SiPM-based camera for the LST telescope of the CTA observatory.
Since our plan was to study only the pixel, the preamplifier and the slow control, we kept most of the existing camera readout.
Referring to Figure~\ref{fig:LST_PMT_01},
only the photo-sensors, their power supply, and the pre-amplifying stage are modified while the rest of the camera would remain unchanged \cite{2019_IEEE_LST_SiPM}.
\par
\par
With our work we wanted to demonstrate some of the main
advantages of SiPMs over PMTs in our application, i.e.
their robustness and tolerance to high illumination levels.
Although SiPMs can withstand currents of several milliamps,
the front-end electronics usually cannot.
Therefore, it is necessary to limit the current flowing into the SiPM. The usual method is to use a protective resistor on the power line. However, this approach has a significant drawback of making the SiPM bias voltage dependent on the current flowing through it, thereby affecting its gain, dark count rate, optical crosstalk, and photon detection efficiency. 
To overcome this problem, we developed an alternative solution using a transistor-based current limiter.
\par
\begin{figure}[!tp]
    \centering
    \subfloat[\label{fig:PIXEL_TO_PCB_M02}]{\includegraphics[width=0.48\columnwidth]{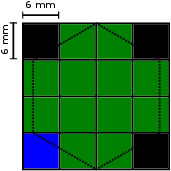}}
    \hfill
    \subfloat[\label{fig:FBK_HD3-4_Array_Picture_01}]{\includegraphics[width=0.48\columnwidth]{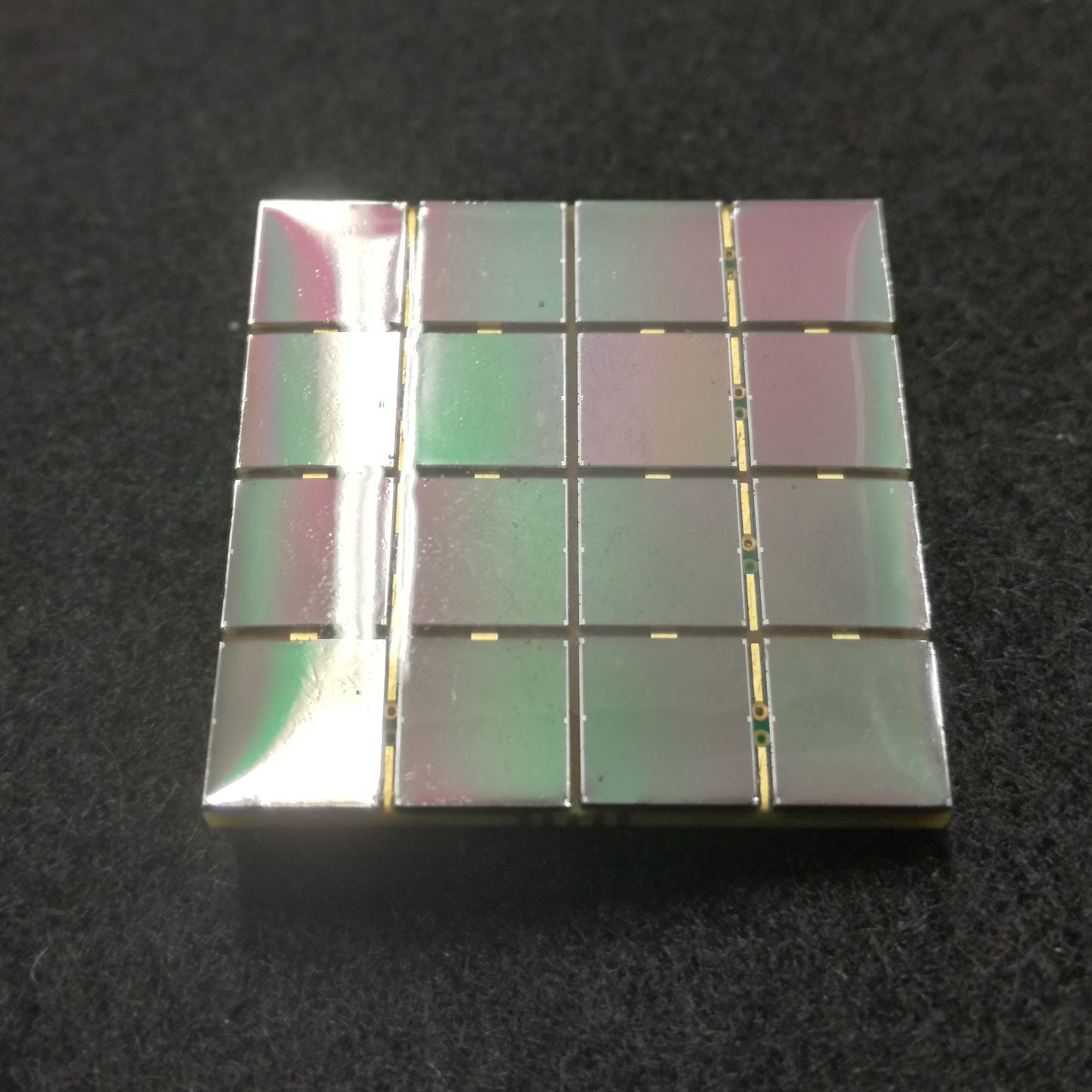}}
    \caption
    {
        (a) SiPM array connection scheme. The signals from the inner twelve SiPMs are summed in the frontend electronics. One of the corner SiPM is used for calibration purposes and disconnected during normal operation. The three remaining corner SiPMs are not used. The dashed line shows the inner area of the Winston cone.
        \\
        (b) A picture of the FBK NUV HD3-4 \qtyproduct{6 x 6}{\milli\metre} $\num{4} \times \num{4}$ SiPM array used for the prototype.
    }
    \label{fig:PIXEL_TO_PCB_M01_02}
\end{figure}

In order to reach sensitive areas comparable to one PMT, the signals of at least twelve \qtyproduct{6 x 6}{\milli\metre} SiPMs need to be summed.
To speed-up the development, we used an already existing $\num{4} \times \num{4}$ SiPM array, connecting only twelve of the sixteen SiPMs. Figure~\ref{fig:PIXEL_TO_PCB_M02} shows which twelve pixels are summed in the frontend electronics  to form one effective camera pixel. One of the corner pixels is used for calibration purposes.
The dashed line shows the inner area of the LST Winston cone, used to concentrate the light onto the photomultipliers and reduce the dead space on the focal plane \cite[]{Okumura_2017}.
The sensors used in this prototype are the NUV HD3-4 from the Fondazione Bruno Kessler (FBK), developed for the mid-size Schwarzschild-Couder Telescope (SCT) of the Cherenkov Telescope Array (CTA) \cite{SCT_SiPM}.
A picture of the SiPM array is shown in Figure \ref{fig:FBK_HD3-4_Array_Picture_01}.
For the final Advanced Camera, we will optimize the matrix layout and the SiPM characteristics.
\par
Given the large number of channels, we picked the Multiple Use SiPM Integrated Circuit (MUSIC), an Application Specific Integrated Circuit (ASIC), to realize the front-end electronics. MUSIC is developed at the Institute of Cosmos Sciencies - University of Barcelona (ICCUB) \cite{MUSIC_SPIE}.
The chip has been specifically developed to readout SiPMs providing 8 channels with an input bandwidth of \qty{500}{\mega\hertz}.
It provides two differential outputs each providing the sum signal of the eight input channels. The two outputs will be connected to the high gain and low gain lines respectively of the LST camera readout.
The output signal can be shaped with an integrated tunable pole zero cancellation circuit, which we configured to yield the shortest signal full width at half maximum (FWHM).
Indeed, unprocessed SiPM signals typically show a long tail due to the recharging of the triggered cells.
With this configuration the peak of a one photoelectron signal corresponds to about 10 ADC channels.
\par
Since the MUSIC ASIC has only eight channels, it is not possible to
directly
sum the signals of twelve SiPMs. 
We solved that problem by grouping the twelve SiPMs into pairs. The signals of one SiPM pair connect in parallel to one MUSIC channel thus using six of the eight MUSIC channels.
\par
\begin{figure}[!tp]
    \centering
    \includegraphics[width=\columnwidth]{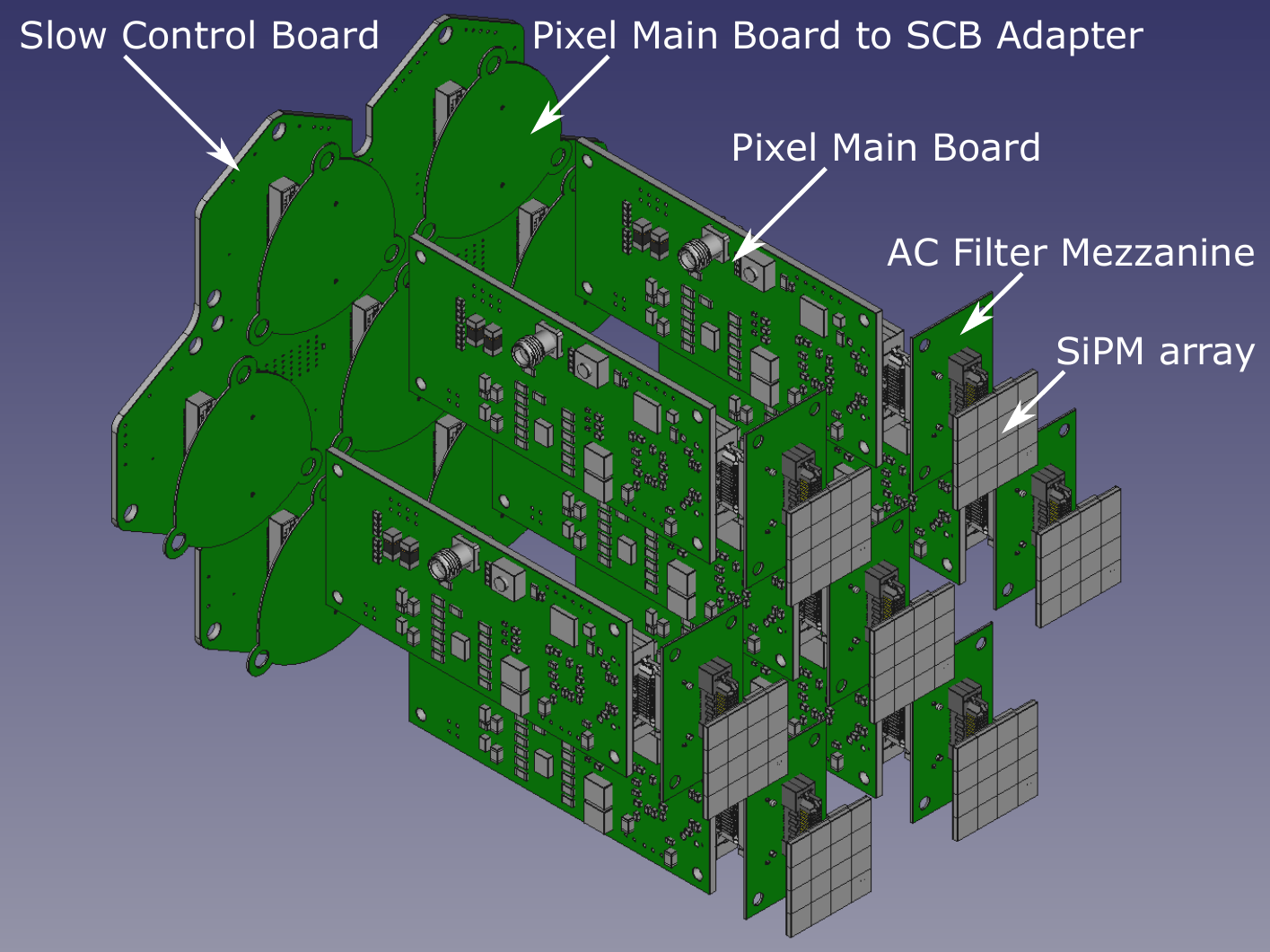}
    \caption{Rendering of the seven-channels module.}
    \label{fig:20211014_3D_Global_View_01_cut_4-3_Names}
\end{figure}

Figure
\ref{fig:20211014_3D_Global_View_01_cut_4-3_Names} shows a rendering of the seven-pixel SiPM module.
Starting from the left, we find the Slow Control Board (SCB), the only part of the LST electronic chain shown here.
Seven SiPM Pixel main boards are connected to the SCB with the help of a printed circuit board (PCB) adapter.
The pixel main board contains the SiPM bias regulator and the front-end electronics.
A mezzanine can be added between the main board and the SiPM array to AC couple the SiPM signals into the front-end. In addition, each channel has a small board with a digital thermometer that is in thermal contact with the back of the SiPM array (here not shown).
\par
\begin{figure}[!tp]
    \centering
    \includegraphics[width=\columnwidth]{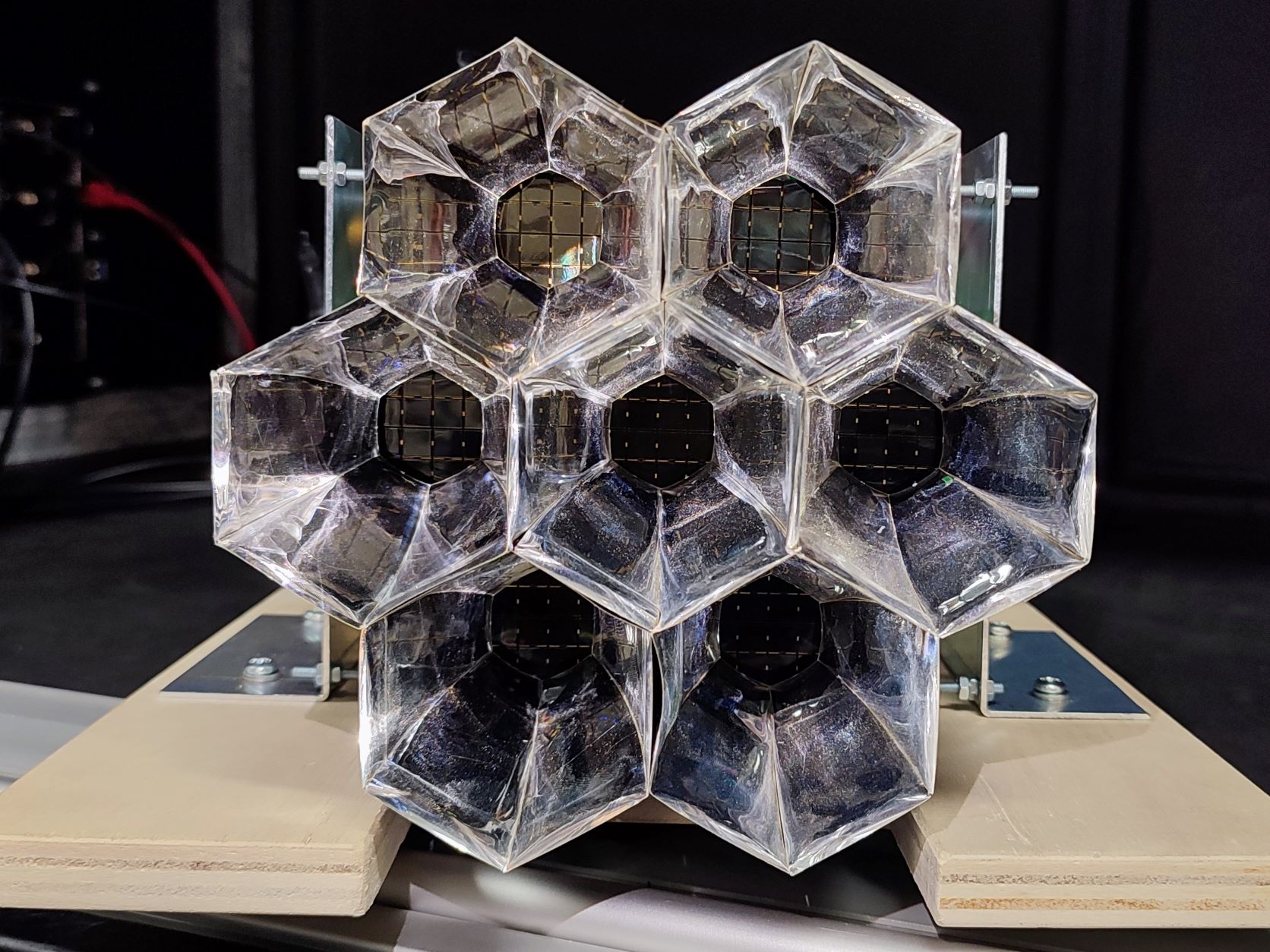}
    \caption{
        Picture of the assembled seven-channel module;
        the mounted Winston cones are preliminary prototypes not used in the current telescope.
    }
    \label{fig:Picture_Global_Module}
\end{figure}

A picture of the assembled seven-channel module is shown in Figure \ref{fig:Picture_Global_Module}.
In the next sections we discuss the design, test and measurements of a single module.

\section{Pixel Design}
%
\subsection{Block Diagram}
\begin{figure}[!tp]
    \centering
    \includegraphics[width=\columnwidth]{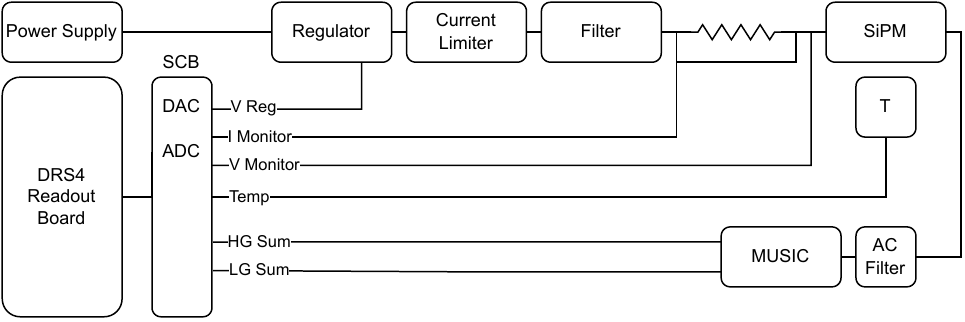}
    \caption{SiPM Pixel Block Diagram.}
    \label{fig:Basic_Scheme}
\end{figure}

The block diagram of the SiPM Pixel is shown in Figure~\ref{fig:Basic_Scheme}.
As already mentioned, this prototype uses most of the LST readout chain, in particular the DRS4 readout board \cite{Masu_DevPMT} and the SCB \cite{SCB}.
\par
A \textit{voltage regulator} (briefly described in section \ref{sec:VoltageRegulator}) is controlled by the \qty[]{12}{bit} Digital to Analog Converter  (DAC) located on the SCB. The regulator output connects a \textit{current limiter} (section \ref{sec:CurrentLimiter}), 
limiting the current flowing into the SiPMs. 
Following the current limiter is a low-pass filter used to filter out noise on the SiPM bias line.
Both the current and the bias voltage are monitored by the Analog to Digital Converter (ADC) on the SCB. The front-end electronics is based on the MUSIC ASIC \cite{MUSIC_SPIE}, which deals with the summing and shaping of the signals.
The regulator, the current limiter, the filter, the voltage and current monitors, and the MUSIC are located on the pixel main board.
In order to be functional even in high background light level, it is possible to insert a high-pass filter in between the SiPMs and the MUSIC, which removes the DC baseline offset.
This filter is located on a dedicated mezzanine, which is removable, to study the difference between DC and AC coupling in our application.
The digital temperature sensor monitoring the SiPM temperature sends its data to the SCB and therefore to the control software. We use the temperature data to adjust the SiPM bias voltage thus correcting for the temperature dependent breakdown voltage of the SiPMs and guaranteeing stable sensor gain.
In principle, this procedure corrects for any temperature increase, including SiPM power dissipation at high count rates.
\par
The SiPMs draw more current than PMTs because they have a higher gain (typically $\sim 10^6$ at operational overvoltage) and because they have a higher photon detection efficiency in the near infrared, where the night sky background is predominant.
On this prototype, with these SiPMs biased at \qty[]{6}{\volt} overvoltage, and without using a UV filter, we expect a current of about \qty[]{1}{\milli\ampere} during observations in dark condition.
Since the SiPMs and front-end electronics require more power than the PMT modules, the SiPMs cannot be powered by the standard LST electronics.
For our prototype we opted for a separate low noise bench power supply.
\par
The DRS4 Readout Board connects to a backplane board, which communicates via Ethernet to a control computer running the LST data acquisition software.
The acquired data are calibrated with the \textit{cta-lstchain} pipeline, developed for the LST data analysis \cite{lstchain_zenodo, lstchain_icrc}.
\par
In the next sections, we will describe the functionalities and the development of each block in our prototype SiPM pixel module.
%
%
\subsection{Voltage Regulator}
\label{sec:VoltageRegulator}
The voltage regulator keeps the SiPM overvoltage constant over a large range of current flowing through the sensor.
Its output voltage is set between \qty[]{0}{\volt} and \qty[]{48}{\volt} by the \qty[]{12}{\bit} DAC located on the SCB.
%
\subsection{Current Limiter}
\label{sec:CurrentLimiter}
The current limiter is used to protect the MUSIC from a damaging high SiPM current.
We demonstrated
that a current of \qty{3}{\milli\ampere} divided over six MUSIC channels would not damage the device, and that it is preferable not to exceed this limit.
The current limit can be relaxed in AC coupling mode, since it prevents the SiPM DC current from flowing into the MUSIC.
Based on Monte Carlo simulations, we expect a current of approximately \qty[]{1}{\milli\ampere} in dark conditions and up to \qty[]{4}{\milli\ampere} at the maximum NSB level at which the LST camera should operate.
This factor already makes AC coupling a preferable option over DC coupling.
\par
The easiest way to limit the current in a SiPM is to use a \textit{protection resistor}.
This approach is good
for a low background detector. However, in applications where the background and thus the SiPM current fluctuates over wide margins like in IACTs, the same approach would result in large changes in bias voltage. 
\par
\begin{figure}[!tp]
    \centering
    \includegraphics[width=0.7\columnwidth]{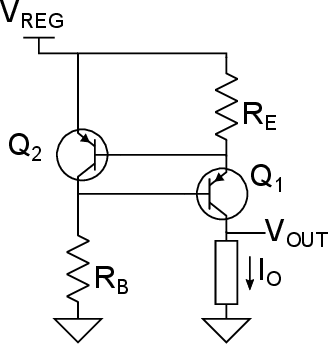}
    \caption{Current limiter.}
    \label{fig:Current_Limiter}
\end{figure}
To limit the current we opted instead for a typical two-BJTs current limiter circuit, shown in Figure~\ref{fig:Current_Limiter}.
That solution results in a stable and basically current-independent bias voltage across the SiPMs. 
\par 
When the current through $R_E$ exceeds the design limit, $Q_2$ turns on, thus turning off $Q_1$, therefore limiting the load current.
If the two transistors have the same characteristics and are mounted nearby in the circuit, they will be at the same temperature, so we can say that they have the same current gain $\beta$ and the same base to emitter voltage $V_{BE}$. 
The current is limited when both $Q_1$ and $Q_2$ are on; the output current $I_O$, which is the current flowing in the load $R_L$, is equal to:
\begin{equation}
    I_{O} = \beta I_{B1} =
    \frac{\beta}{\beta+1} \left(
    I_{B2} + I_{R_E}
    \right) =
    I_{MAX}
\end{equation}
With simple calculations we obtain:
\begin{equation} \label{eqn:I_MAX}
    I_{MAX}
    = \frac{\beta}{(\beta + 1)\beta + 1} \left(
    \frac{V_{BE}}{R_E} \left(
    -\beta
    \right) +
    \frac{V_{REG} + 2 V_{BE}}{R_{B}}
    \right)
\end{equation}
In DC mode we want to limit the current to \qty{3}{\milli\ampere}, so we chose $R_B = \qty{10}{\kilo\ohm}$ and $R_E = \qty{220}{\ohm}$.
When operating in AC mode, the  current limit can be higher (we choose \qty{10}{\milli\ampere}) and it is set adding  a \qty{100}{\ohm} resistor in parallel to $R_{E}$.
\par
For the actual circuit we had to choose a transistor with a collector-emitter breakdown voltage $V_{(BR)CEO}$ above the SiPM bias voltage: we choose the \textit{MMBT5401L}, whose minimum $V_{(BR)CEO}$ is \qty{-150}{\volt}.
%
\subsection{Voltage and Current Monitors}
The regulator output voltage and current are monitored by a voltage and a current monitor.
The voltage monitor is an inverting amplifier, which, as sketched in Figure \ref{fig:Basic_Scheme}, reads the voltage directly from the SiPM cathode.
The current monitor is a differential amplifier, which reads the voltage drop on a resistor placed between the Filter and the SiPM array.
The amplifier outputs are connected to the ADC on the SCB.
%
\subsection{MUSIC and Microcontroller}
The MUSIC ASIC is controlled by an ATmega328P-AU microcontroller which runs the MUSIC MiniBoard firmware \cite{MUSIC_MiniBoard}.
\par
To ensure that the MUSIC bias is in its operational limits, two Low Dropout Regulators are mounted on the pixel main board, one for the \qty{3.3}{\volt} line and one for the \qty{5.0}{\volt} one.
%
\subsection{High-Pass Filter}
%
In high background applications, the signal lies over a quasi-DC noise.
This DC baseline can be removed with an high-pass filter
between the SiPM anode and the MUSIC input; we refer to this configuration as AC coupling.
However, by adding this filter, we lose MUSIC's ability to individually adjust the SiPM bias voltages to compensate for the breakdown voltage difference in the array. The effect of the possible gain differences should be evaluated by further studies.
\par
If $R_F$ is the filter resistor, the voltage across the SiPM $V_{SiPM}$ would be, considering only the DC component:
\begin{equation} \label{eqn:AC_VSiPM}
    V_{SiPM} = V_{bias} - I_{SiPM} R_F
\end{equation}
where $V_{bias}$ is the SiPM bias voltage and $I_{SiPM}$ is the current flowing into the sensor.
Therefore, to keep constant the voltage across the SiPM, we need to actively adjust the bias voltage $V_{bias}$ with a proper feedback loop, in order to compensate for the voltage drop on $R_F$.
If the value of the $R_F$ resistor is too high, a significant voltage drop will occur across it. Conversely, if the value of the resistor is too low, the signal loss would be too high because the resistor is in parallel with the MUSIC input. We chose $ R_F = \qty[]{100}{\ohm}$.
\par
Regarding the filter capacitor $C_F$, we found in measurements that a \qty{1}{\micro\farad}  AC coupling capacitor does not change the signal shape and, so we chose this value.
\par
Each of the six channels contributing to the sum has the \textit{R-C} filter previously described; the calibration pixel is instead directly connected, with no filter on its line.
\par
\subsection{Digital Thermometer}
Since most of the SiPM characteristics depend on the overvoltage and due to the high temperature dependence of the breakdown voltage $V_{bd}$, 
it is necessary to precisely monitor the sensor temperature, in order to properly correct the bias voltage.
For example, for the NUV HD3-4 we measured a $d V_{bd} / d T $ equal to \qty[per-mode=symbol]{27}{\milli\volt\per\celsius}.
A \qtyproduct{1 x 1}{\centi\metre}  PCB with the digital thermometer is attached with thermal paste to the back of the SiPM array. The temperature values are sent to the LST electronics and from their recorded by the slow control software.
\subsection{Design of the Pixel Main Board}
%
\begin{figure}[!b]
    \centering
    \includegraphics[width=\columnwidth]{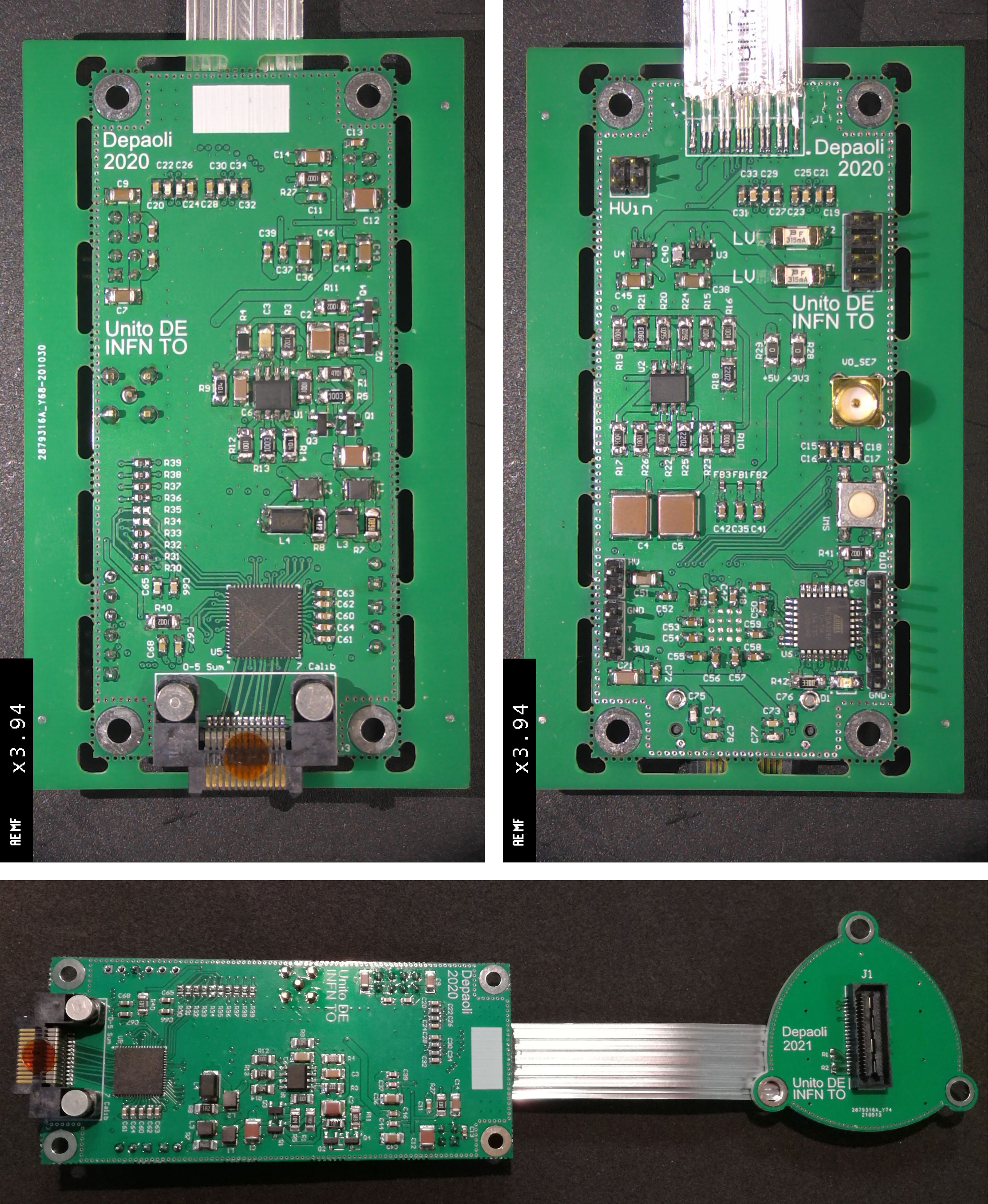}
    \caption{
        Top: pictures of the pixel main board. Bottom: pixel main board connected to the Slow Control Board Adapter PCB.
    }
    \label{fig:Pixel_TO_Main_Board_Photo_03}
\end{figure}
The design started with the MUSIC MiniBoard schematics and PCB layout, kindly made available by the MUSIC developers which made it possible to use the MiniBoard firmware in our design.
The main board is a six-layer PCB, with the two inner layers dedicated only to the high-speed sum signals (one for the High Gain, the other one for the Low Gain).
Most of the components were assembled at Asseltech Srl, Chivasso, Piedmont, Italy, and the remaining part at the Electronics Laboratory of the
National Institute for Nuclear Physics (INFN), Turin section.
Pictures of the assembled board are shown in Figure~\ref{fig:Pixel_TO_Main_Board_Photo_03}.
%
\subsection{Camera Control Software}
%
\begin{figure}[!tp]
    \centering
    \includegraphics[width=\columnwidth]{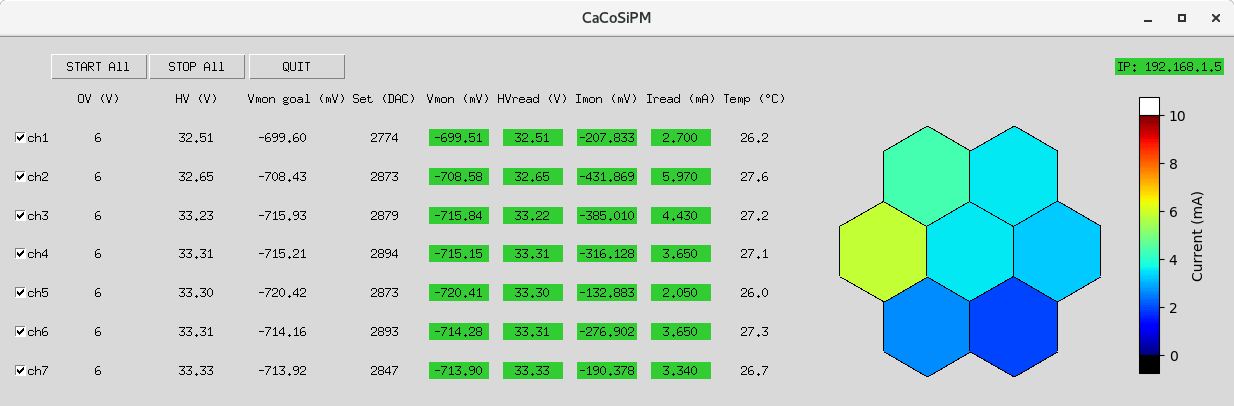}
    \caption{CaCoSiPM user interface.}
    \label{fig:CaCoSiPM_Interface_Module}
\end{figure}
The Camera Control Software for the SiPM Pixel (\textit{CaCoSiPM})
is the software that takes care of the module slow control.
It runs on a computer connected via Ethernet to the DRS4 readout board and is based on the software developed by the LST collaboration to control the LST modules.
The python-based GUI designed to control the system is shown in Figure~\ref{fig:CaCoSiPM_Interface_Module}.
\subsubsection{Normal Operation}
At start-up, the SiPM voltage is set above zero, but below the
breakdown to keep it from producing signals but still have it reverse biased. This is done
to avoid damaging the MUSIC with the high current that could ﬂow
through the SiPM junction if forward biased.
\par
The software will activate the channel, adjust the SiPM bias voltage to the desired level, and maintain it through a feedback loop operating at \qty[]{1}{\hertz}.
The feedback works in the following way: the software evaluates the required $V_{HV}$ corresponding to the desired overvoltage, then calculates the corresponding voltage monitor $V_{mon}$ value and changes the DAC value to reach the desired $V_{mon}$ (one DAC bit corresponds to about \qty[]{8}{\milli\volt}  SiPM bias voltage).
The regulator output voltage $V_{HV}$ is calculated as follows:
\begin{equation}
    \label{eqn:V_HV}
    V_{HV} = V_{OV} + V_{bd}(T_0) + \frac{d V_{bd}}{d T} (T - T_0) + V_{anode}
\end{equation}
Where $V_{anode} = V_{MUSIC IN}$ in the DC and $V_{anode} = R_{AC} * I_{mon}$ in the AC coupling.
$V_{OV}$ is the desired SiPM overvoltage at which we want to operate the pixel, \qty{6}{\volt} in our prototype.
$V_{bd}(T_0)$ is the average breakdown voltage of the twelve SiPMs in the array forming the camera pixel at the reference temperature $T_0$.
The third term is the temperature compensation of the SiPM breakdown voltage, which increases by
$d V_{bd}/d T$.
The last correction $V_{anode}$ depends on the SiPM coupling (AC or DC)
and is introduced since in both cases the SiPM anode voltage is higher than zero.
In the DC case, we have seen that the SiPM anode voltage is equal to the MUSIC input pin voltage $V_{MUSIC \, IN}$.
In the AC case the situation is instead a bit more complicated. The DC current flowing through the SiPM causes a voltage drop across the filter resistor:
\begin{equation}
    V_{R_F} = I_{SiPM} R_F
\end{equation}
where $I_{SiPM}$ is the current flowing through the SiPM. By design, the current read by the Current Monitor $I_{mon}$ is the sum of the currents flowing into the twelve SiPMs; since every two SiPMs are connected in parallel, we have six channels in total. With good approximation, the current splits equally between the six channels, and thus $V_{anode}$ will be:
\begin{equation}
    V_{anode} = \frac{I_{mon}}{6} R_F
\end{equation}
\par
The software evaluates therefore the value of $V_{HV}$ and, through a feedback loop, adjusts the DAC value accordingly to achieve the desired bias voltage for the SiPM.
%
\section{SiPM Pixel Measurements}
The seven-pixel module is kept in a light-tight dark box.
The CaCoSiPM software controls and monitors the SiPM bias voltage; all the measured values are saved in a log file.
A DC LED controlled with a power supply simulates the night sky background;
by increasing the LED bias it is possible to scan different NSB-equivalent intensities, corresponding to different currents flowing into the SiPM array.
The sensor is also illuminated with a pulsed sub-nanosecond LED (Picoquant PDL 800-D with PLS 400 LED head), simulating a fast Cherenkov flash.
\subsection{Comparison Between AC and DC Coupling}
\begin{figure}[!hbtp]
    \centering
    \includegraphics[width=\columnwidth]{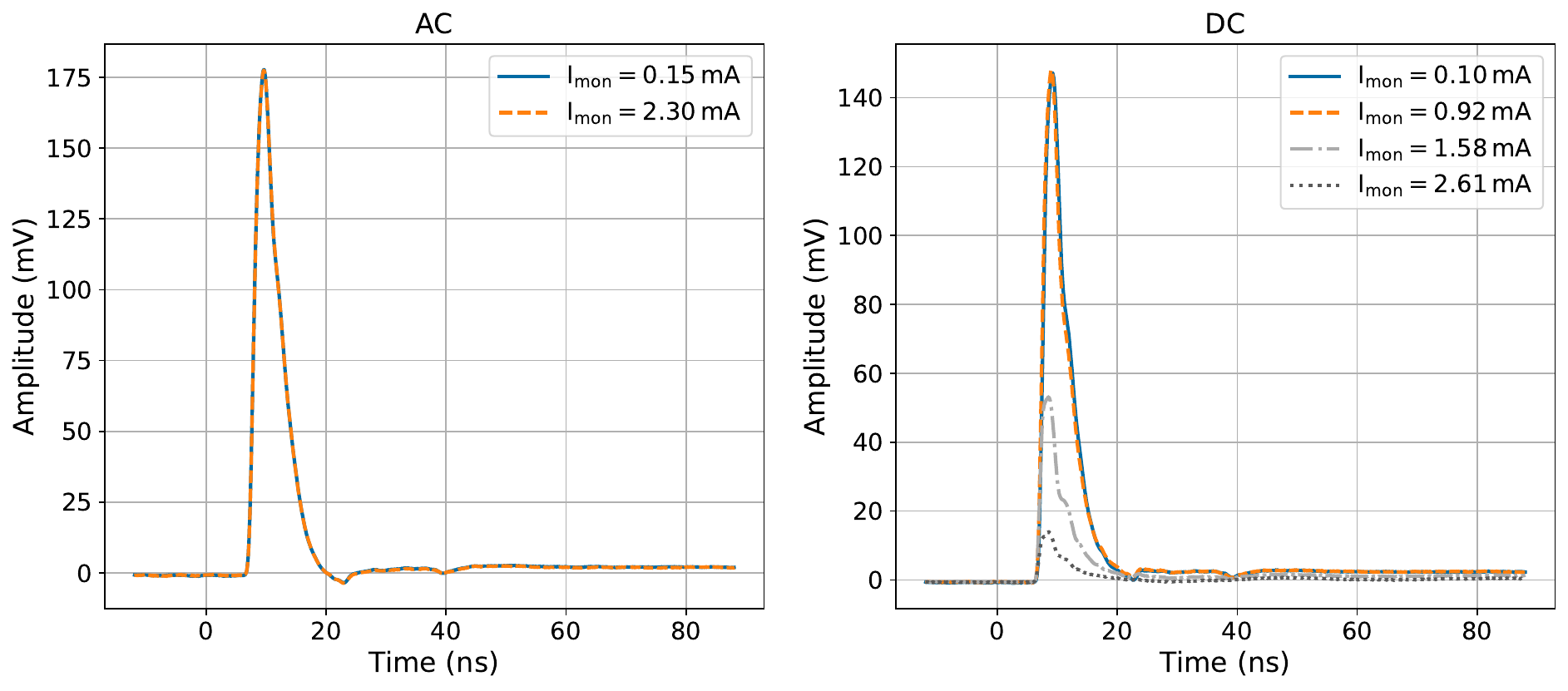}
    \caption{
        Signals acquired in AC (left) and DC (right) coupling at different light backgrounds;
        in both plots the solid and dashed waveforms are overlapping.
    }
    \label{fig:20210225_AC_vs_DC_01}
\end{figure}

From laboratory measurements, 
shown in Figure \ref{fig:20210225_AC_vs_DC_01}, where on the left we have the High Gain signals acquired in AC coupling, and on the right in DC coupling,
we observed that in the DC approach the MUSIC output drops as soon as the combined current of all SiPMs connected to the same MUSIC exceeds \qty{1}{\milli\ampere}.
Signals are acquired, thanks to a custom adapter, by means of a Tektronix MSO6 Oscilloscope with \qty{1}{\giga\hertz} bandwidth and with a \qty{100}{\nano\second} acquisition window.
Each waveform is the average of \num{1000} acquisitions, and is labelled with the corresponding current flowing in the channel, measured with the Current Monitor.
Thus, the AC coupling is the best solution for our application, where working in a high-brightness background environment is a critical aspect.
\par
In addition, the current limiter hardware settings can be relaxed if we work in AC mode, since the DC current due to the night sky background would flow to the ground through the filter resistor.
\par
From now on all the reported measurements are done
in AC coupling mode
and with the current limiter set to \qty[]{10}{\milli\ampere}.
%
%
\subsection{Current Limiter Test}
\begin{figure}[!btp]
    \centering
    \includegraphics[width=\columnwidth]{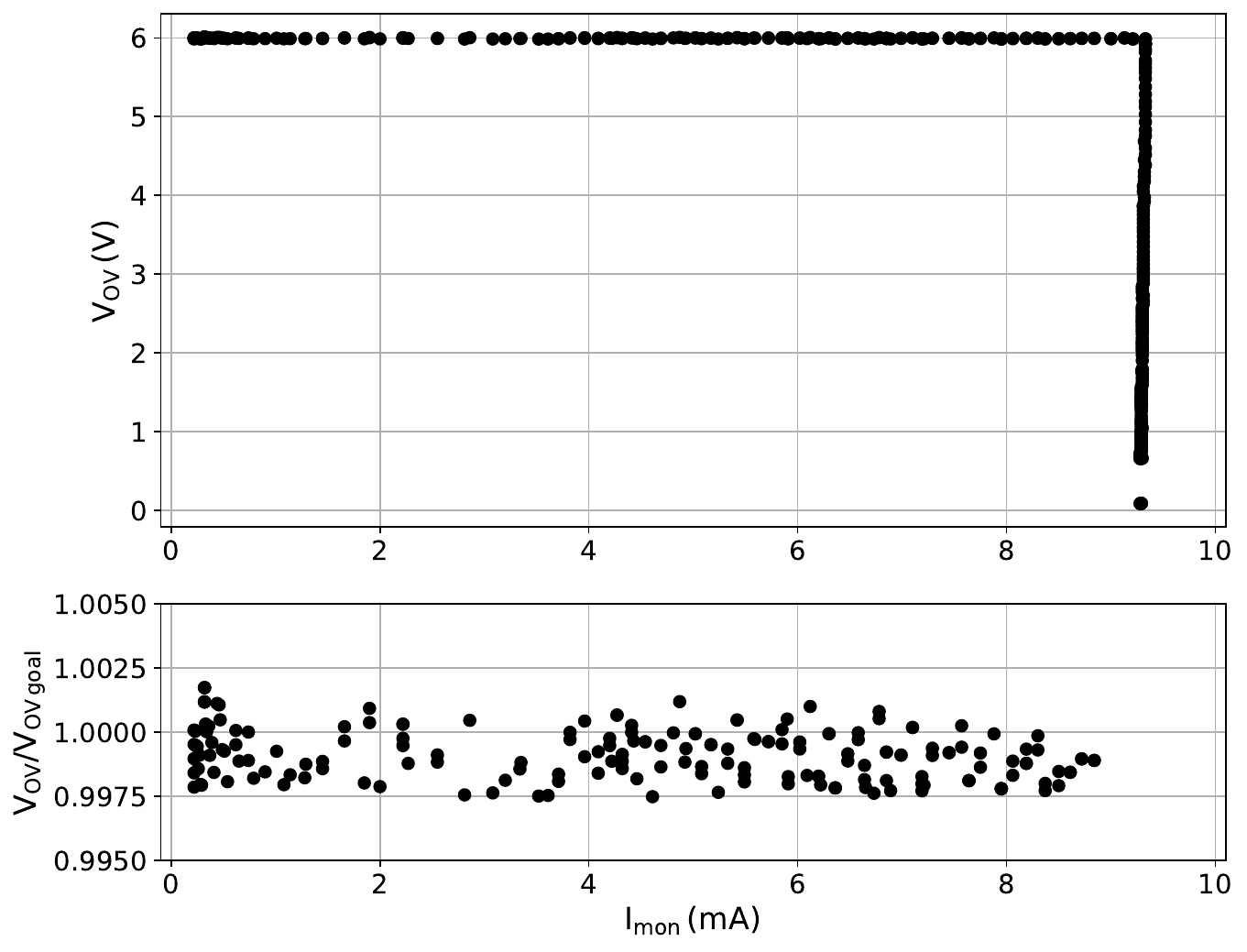}
    \caption
    {
    Top: SiPM overvoltage dependence on current. Bottom: ratio of measured overvoltage and desired one before current limit intervention.
    }
    \label{fig:20210714_162559_CaCoSiPM_OV_Norm}
\end{figure}

The results of the current limiter tests are shown in Figure \ref{fig:20210714_162559_CaCoSiPM_OV_Norm} for a single channel, where $I_{mon}$ is the current flowing in the SiPMs and $V_{OV}$ is the SiPM overvoltage.
The datapoints at the lowest value for $I_{mon}$ are obtained by switching the LED off. The non-zero $I_{mon}$ current is due to SiPM intrinsic dark counts.
Turning on the LED and gradually increasing its brightness, we can see how the SiPM overvoltage remains constant within about \qty[]{0.25}{\%}, until it drops sharply when the hardware-set limit is reached. The last points, the ones at the lowest $V_{OV}$ values, are taken with the dark box open, exposing the switched-on detector to daylight proving its robustness.
\par
%
\subsection{Waveform acquisition}
%
\begin{figure}[!hbtp]
    \centering
    \includegraphics[width=\columnwidth]{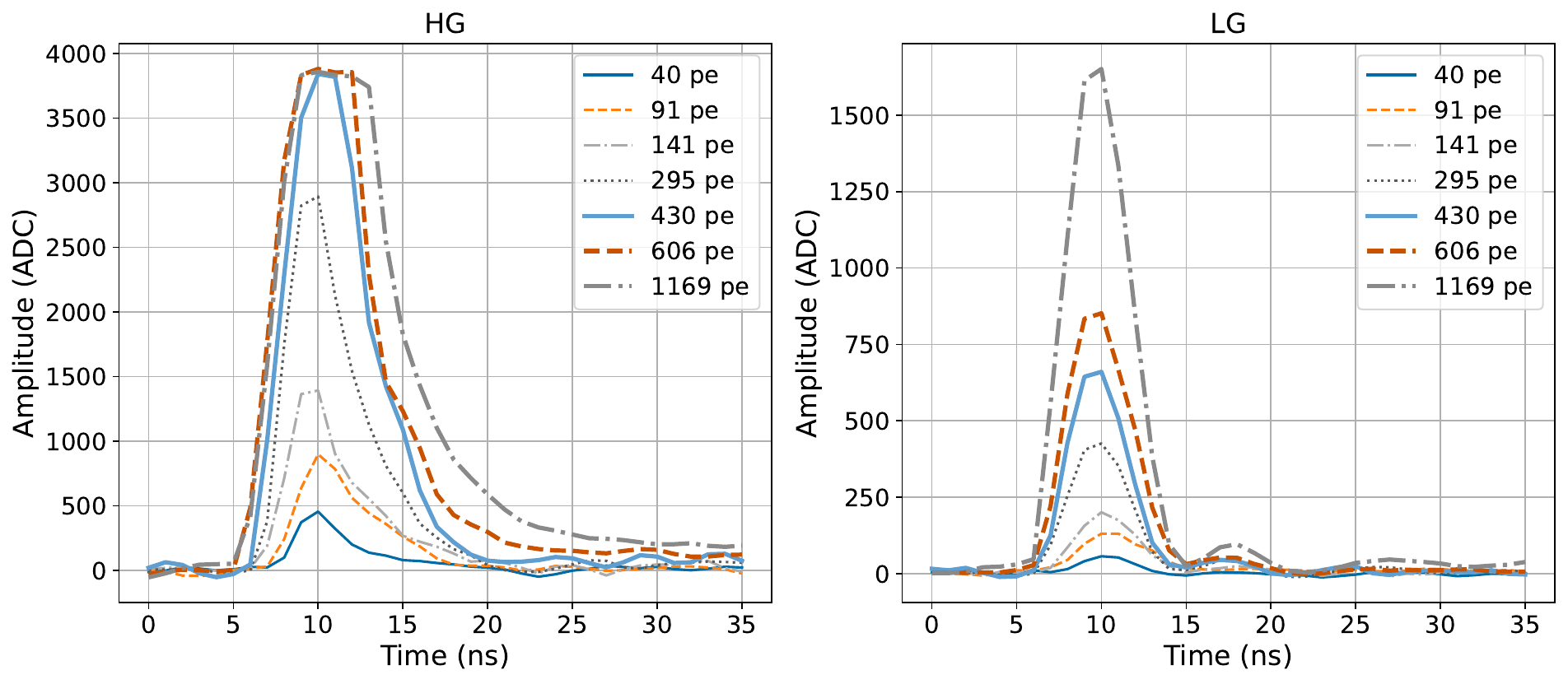}
    \caption{
        Examples of High Gain (left) and Low Gain (right) SiPM Pixel calibrated waveforms acquired with the LST electronics;
        waveforms of the same line style refer to the acquisition of the same light pulse intensity.
    }
    \label{fig:20210708_Waveforms_HG_LG_02}
\end{figure}
The SiPM Pixel signal acquisition is based on the LST readout electronics.
The
High Gain and Low Gain signals are acquired with the control computer connected via Ethernet to the DRS4 readout board.
The signal processing relies on the \textit{cta-lstchain} software package.
Examples of High Gain (a) and Low Gain (b) waveforms at different pulsed LED intensities are shown in Figure
\ref{fig:20210708_Waveforms_HG_LG_02}.
Waveforms of the same line style refer to the acquisition of the same photon pulse intensity;
the number of photoelectrons reported in the labels is obtained from the calibration factors that will be discussed later.
\par
Although the High Gain signal has a longer tail than the Low Gain one due to the different amplifiers in the two channels on the LST electronics board, this has no effect on the measured FWHMs:
\begin{equation}
    FWHM_{HG} \simeq FWHM_{LG} \simeq \qty{4.3}{\nano\second}
\end{equation}
\par
This result is similar to what we measured for the PMT (\qty{\sim 3}{\nano\second} for the High Gain and \qty{\sim 3.5}{\nano\second} for the Low Gain), so we expect similar performance, which should be confirmed by a detailed Monte Carlo simulation.
%
\subsection{Pixel Calibration}
The twelve SiPMs constituting one camera pixel are connected two by two in parallel to six MUSIC ASIC channels, which sums and shapes their signals.
It is impossible to distinguish single photoelectron signals in the sum signal, neither in High Gain nor Low Gain. It is therefore necessary to find another solution to calibrate the pixel.
We connected one of the unused MUSIC channels to one of the four SiPMs at the corners of the array, normally disconnected,
since we are able to see single photoelectron signals on this channel and we can use it to cross-calibrate the whole pixel.
The signal from this SiPM can be accessed from the corresponding single-ended (SE) MUSIC output and it is not shaped by the pole-zero cancellation circuit, thus preserving its amplitude. The Differential Leading Edge Discriminator (DLED) technique \cite{DLED} is used to deal with the long SiPM tails in the recorded signal traces.
\par
The SE signal from the reference pixel
is digitized with an oscilloscope.
\par
During the calibration procedure, the SiPM pixel (without the Winston cones) is illuminated with a uniform pulsed light. The light uniformity is obtained by means of an optical diffuser, and has been evaluated to be on the order of \qty{7}{\%}.
The pulsed LED triggers the acquisition of the SE reference signal and the High Gain and Low Gain sum signals.
In this configuration, if the reference pixel is hit by an average number of photons $n_{ph,SE}$, the twelve-SiPM pixel, simply by scaling to the different effective sensor areas, will detect on average twelve times more photons.
With good approximation, all SiPMs in the 4 x 4 array have the same photon detection efficiency. Therefore, the number of detected photons, i.e. photoelectrons, is also twelve times higher in the combined twelve-SiPMs as compared to the SE SiPM.
Therefore, by knowing the average number of photoelectrons generated in the reference SiPM, we can calibrate the whole pixel.
If $A_{HG}$ and $A_{LG}$ are the average area of the High Gain and Low Gain sum signals respectively, the HG and LG gains are defined as:
\begin{equation} \label{eqn:Calibration_gain}
    \begin{split}
        g_{HG} & = \frac{A_{HG}}{n_{pe,SUM}} \simeq \frac{A_{HG}}{\num{12} \, n_{pe,SE}}
        \\
        g_{LG} & = \frac{A_{LG}}{n_{pe,SUM}} \simeq \frac{A_{LG}}{\num{12} \, n_{pe,SE}}
    \end{split}
\end{equation}
\par
As a first step, we checked that the amplitudes of the sum signals do not change when going from operation to calibration mode.
Then the reference pixel is calibrated and its linearity is studied.
\par
\begin{figure}[!t]
    \centering
    \includegraphics[width=\columnwidth]{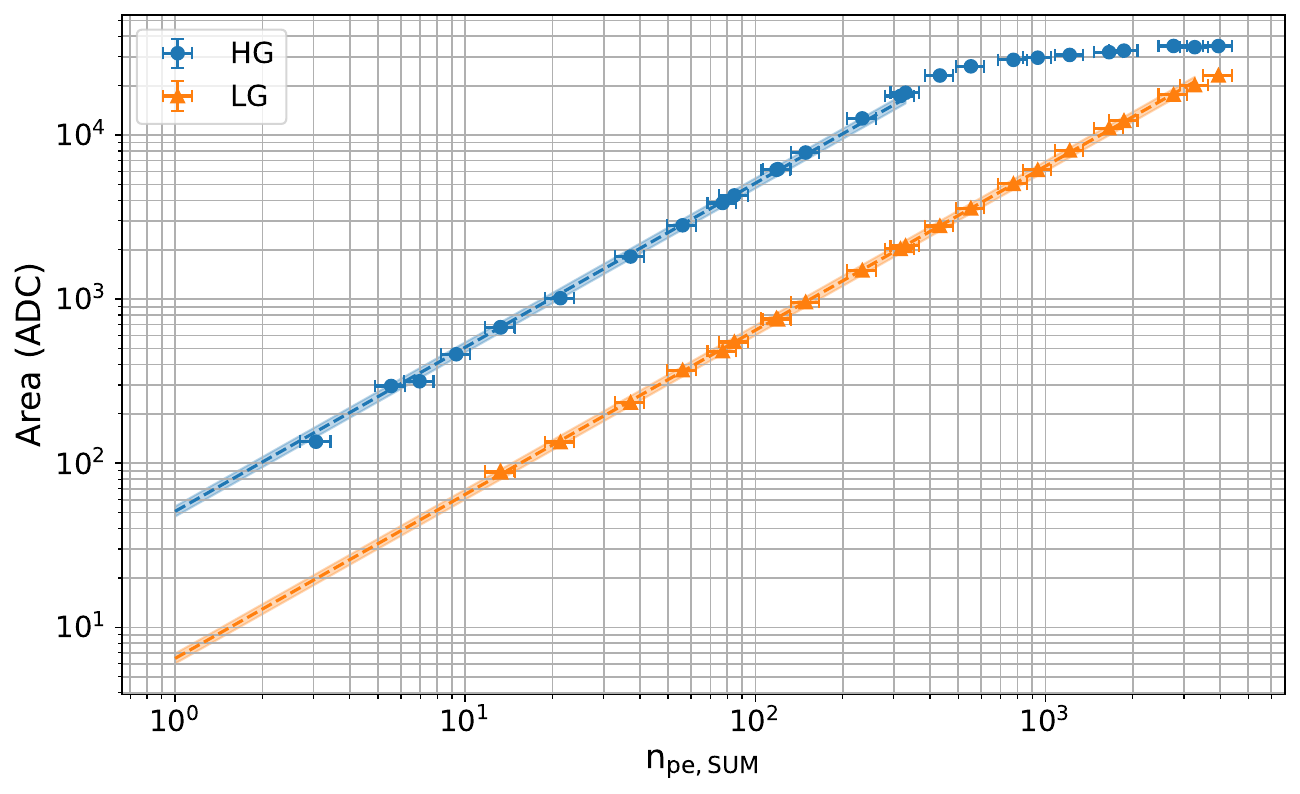}
    \caption
    {
        SiPM pixel calibration and linearity results.
        The error bands take into account the statistical error and the light non-uniformity.
    }
    \label{fig:20210708_Calibration_SUM_Article}
\end{figure}

The calibration procedure is repeated for different pulsed LED light intensity, leading to the results shown in Figure~\ref{fig:20210708_Calibration_SUM_Article}.
The High Gain channel saturates at about 300 photoelectrons, while the Low Gain is linear up to 2000 photoelectrons.
Data in the linear region are fitted with:
\begin{equation}
    \begin{split}
        A_{HG} & = g_{HG} \, n_{pe,SUM} \\
        A_{LG} & = g_{LG} \, n_{pe,SUM}
    \end{split}
\end{equation}
where $A_{HG}$ and $A_{LG}$ are the integrated High Gain and Low Gain SUM signals and $n_{pe,SUM}$ the number of photoelectrons.
Once $g_{HG}$ and $g_{LG}$ are found,
it is possible to convert every signal into number of photoelectrons.
\par
Further studies would be needed to estimate all systematics effects and the time evolution of the calibration factors in case the module is mounted on the telescope.
%
\section{Conclusions}
In this work we developed and built an SiPM-based prototype module for the Large-Sized Telescopes of the Cherenkov Telescope Array.
In our development we focused on demonstrating the advantages of SiPMs over PMTs in our application,
i.e. their robustness and tolerance to high illumination levels.
In principle, with these sensors it would be possible to observe under illumination conditions that are too bright for traditional photomultipliers, thus leading to an increase in the duty cycle of the experiment.
In addition, their robustness would improve the reliability of the telescope: indeed, accidentally exposing the PMTs to daylight, even with their high voltage off, can seriously damage them.
In particular, using an active current limiter instead of a protection resistor proved to be essential to maintain the same overvoltage at different night sky background levels, and the AC coupling approach, reducing the current flowing through the ASIC, allows us to see signals even in high background conditions.
Using the MUSIC's shaping circuit, we obtained a fast signal for the SiPM module, similar to the PMT one.
The solutions described in this work can be used in a future LST SiPM-based camera, or more in general in any other SiPM applications characterized by a high background noise.
%
\section*{Acknowledgments}
We are grateful for financial support from organizations and agencies listed in \url{https://www.cta-observatory.org/consortium_acknowledgments/} and from ``Dipartimento di Eccellenza, Università degli Studi di Torino''.

\bibliography{Bibliography}

\end{document}